\documentclass[preprint]{aastex} 

\begin{document}

\title{A Survey for EHB Stars in the Galactic Bulge} 

\author{Donald M. Terndrup, Deokkeun An, 
Angela Hansen} 
\affil{Department of Astronomy, The Ohio State University,
Columbus OH 43210, USA} 

\author{Ruth C. Peterson} 
\affil{Astrophysical Advances, Palo Alto CA, USA}

\author{Alistair R. Walker}
\affil{NOAO/CTIO, La Serena, Chile} 

\and

\author{Elaine M. Sadler}
\affil{School of Physics, University of Sydney, NSW 2006, Australia}

\begin{abstract}
We present a progress report on an extensive survey to
find and characterize all types of blue horizontal-branch stars
in the nuclear bulge of the Galaxy.  We have obtained
wide, shallow imaging in $UBV$ of $\approx 12$ square 
degrees in the bulge, with follow-up
spectroscopy for radial velocities and metal abundance
determinations.  We have discovered a number of metal-rich
blue HB stars, whose presence in the bulge is expected by 
the interpretation of the extragalactic ultraviolet excess.  
Very deep images have been obtained in $UBV$ and SDSS $u$ 
along the bulge minor axis, which reveal a significant
number of EHB candidates fainter than $B =19$, i.e., 
with the same absolute magnitudes as EHB stars in several 
globular clusters. 

\end{abstract}

\section{Introduction:  EHB stars and the bulge}  
This meeting is mainly concerned with very hot stars 
on the extreme or extended horizontal branch (EHB).
It is important, however, to place these stars in 
the wider context of the formation and evolution of 
the entire blue horizontal branch (BHB), here meaning 
HB stars that are hotter than the RR Lyrae instability strip.  

As we will hear at this meeting, EHB stars are common
in the local field, and have been discovered in several 
globular clusters and a few open clusters.  Among the 
globulars, only those with metallicity $\leq 0.05$ solar 
(${\rm [Fe/H]} \leq -1.3$) have hot BHB stars in abundance. 
In contrast, very few are found in intermediate-metallicity 
systems such as 47 Tucanae, but they have been discovered in 
small numbers in two metal-rich globulars \citep{rich97,sosin97}.  

The well-known phenomenon of the ultraviolet (UV) upturn 
\citep{cw79,ber80,otp86,dbj02} in massive ellipticals
and the bulges of other spirals tells us that old, metal-rich 
stellar populations can make significant numbers of hot stars 
(see other contributions to this conference). The generally 
accepted sources of the UV light are the EHB stars at the very 
hot end of the horizontal branch, the subdwarf B (sdB) stars 
and their shorter-lived progeny \citep{ocon99}. The numbers 
or lifetimes of EHB stars must increase with metallicity, 
since the far-UV flux does \citep{faber83,bur88,longo89}.   

Locally, metal-rich BHB stars do exist. In the old open 
cluster NGC~6791, \citet{lsg94} find sdB stars, and 
\citet{pg98} confirm a proper-motion and radial velocity 
member as a metal-rich (${\rm [Fe/H]} =+0.4 \pm 0.1$) star
on the cool BHB.  But even in this massive open cluster, 
the EHB is too sparse to 
determine many details of the evolution of these stars: the lifetimes 
of these evolved stars are too short.  

Globular clusters have a different
relationship between metallicity and UV light from
bulges or elliptical galaxies, as illustrated
in Figure 1.   This plots the strength of the UV excess 
(from the 1550 \AA $- V$ color) against metallicity (from 
the ${\rm Mg}_2$ index of Faber et al.\ 1985).   The open points
are for Galactic globular clusters  \citep{dor95}; the one with
the lowest far-UV flux is 47 Tucanae.  The triangles show 
ellipticals and  spiral bulges from \citet{bur88}.  The filled 
triangles on Figure 1 are, in order of increasing ${\rm Mg}_2$, 
for M~32 and M~31. At present, there is no direct measure of 
the 1550 \AA\ flux in the bulge because of the high interstellar 
extinction at low galactic latitudes, but there is  a measurement 
of the integrated ${\rm Mg}_2$ index in the Baade's Window field 
of the bulge, 500 pc from the galactic center \citep{idiart96}. 
This is shown as a vertical line in Figure 1, and indicates the 
range of UV fluxes expected from the inner bulge.  
Perhaps not surprisingly given the luminosity of the bulge, 
we can then expect that the bulge will contain more EHB stars per 
volume than M~32, where EHB stars are known to exist (Brown, 
this conference), but fewer than in M~31.  

\begin{figure}[ht]  
\centerline{\includegraphics[width=23pc]{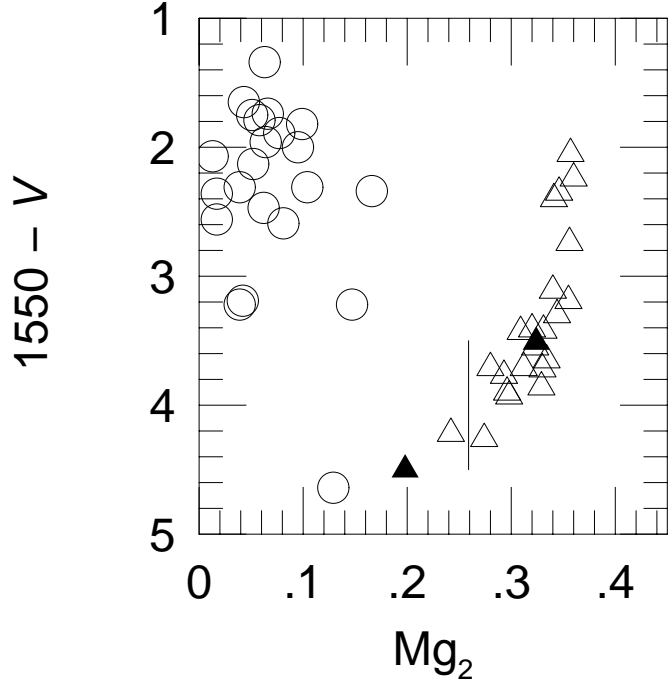}}  
\caption{Strength of UV emission vs.\ the Mg$_2$ metallicity 
index in globular clusters of the Galaxy (open circles), 
and in ellipticals and spiral bulges (triangles).  The 
filled triangles are for M~32 (left) and M~31 (right).
The vertical line shows the value of Mg$_2$ measured 
for the bulge.}
\end{figure}  

Furthermore, the bulge is a good example of an old, metal-rich 
population akin in density and star formation history to that in 
elliptical galaxies and the bulges of other spirals \citep{fw87,fro90}.
Abundances in the inner bulge span [Fe/H] $= -1.0$ to $+0.3$, 
with many stars showing enhanced light-element abundances
\citep{wr83,mcwr94,castro96,srt96}.  This resembles models of 
elliptical galaxy formation \citep{ya87} and interpretations of the
integrated optical spectrum of the central regions of external 
galaxies \citep{w92,wfg92}.   The radial metallicity gradient 
in the bulge \citep{tern88,tfw90,tft95}  provides a natural 
testbed of how metallicity drives BHB formation. The bulge 
is massive enough to support a large BHB population, 
perhaps including the short-lived progeny of EHBs.  Stars of 
all types can be counted and analyzed, allowing a direct determination 
of the contribution of BHB and EHB stars to the UV fluxes of the 
centers of galaxies;  hot HB stars can seriously 
affect the determination of ages and metallicities
deduced for these systems from their integrated optical 
or UV spectra \citep{ponder98,peterson03}.

\section{Survey strategy}  
We have been conducting a survey which will yield
complete samples of BHB stars of all temperatures. 
There are several components to this survey:  

\noindent\hangindent=\parindent\hangafter=0
1) Wide-field, shallow imaging with the CTIO Schmidt 
telescope to identify BHB stars down to a limiting magnitude 
of $B \approx 19$. This is not sufficient to find the EHB 
stars that would create the UV flux (below), but it does 
generate large samples of cooler ($T_{\rm eff} \leq 15000$ K)
BHB stars for follow-up 
spectroscopy to determine metallicities. 
In addition, we have time-series photometry with the Schmidt on 
several lines of sight to identify RR Lyraes, 
whose large excursions in temperature (or non-simultaneous
$UBV$ photometry in the survey) could land them 
in our samples of BHB stars.  

\noindent\hangindent=\parindent\hangafter=0
2) Fiber spectroscopy of BHB stars 
selected from the Schmidt survey.  Stellar temperatures, 
gravities, and abundances are derived from spectral synthesis.
Reddening is found by comparing observed colors with those
predicted for the resulting stellar model. Distances follow
from the apparent magnitude, providing with the radial
velocities a check of membership in the bulge.
A particularly important metallicity indicator is 
the Mg II $\lambda 4481$ \AA\ line, which is not strongly
affected by radiative levitation for $T_{\rm eff} \leq 16000$ K
\citep{behr99}.  Most evolutionary models (e.g., Yi
et al. 1997, 1998)  predict that mechanisms of producing EHB stars at
high metallicity will also cooler ($T_{\rm eff} \sim 10000$ K)
metal-rich BHB stars as well.

\noindent\hangindent=\parindent\hangafter=0
3) Deep imaging in $UBV$ and SDSS $u$ along the 
bulge's minor-axis with the CTIO Mosaic 
camera on the Blanco 4m telescope, to identify and determine
the spatial distribution of faint EHB candidates.  

The lines of sight in our survey are illustrated in Figure 2.  

\begin{figure}[ht]  
\centerline{\includegraphics[width=23pc]{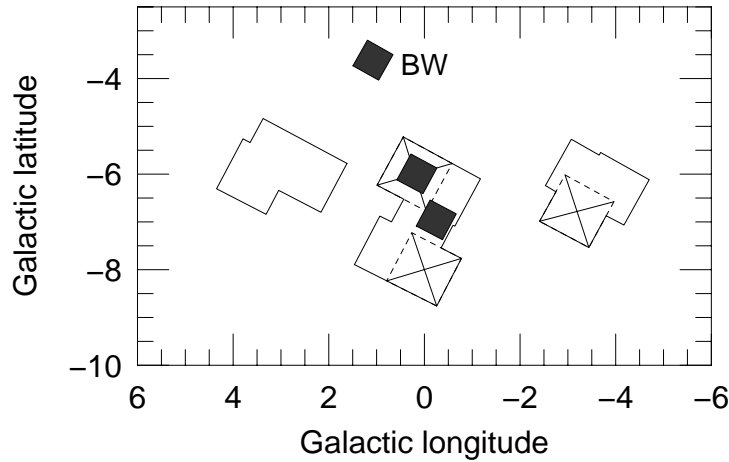}}
\caption{Location of survey regions in the bulge.  The 
polygons outline the areas surveyed with the CTIO Schmidt for 
the cooler, brighter BHB stars.  The squares marked 
$\times$ indicate the fields with AAT spectra in 2001;  
the off-axis one was also observed in 1997 (Peterson
et al., 2001).  
The shaded smaller squares mark fields with the Mosaic 
imaging from CTIO in 2001. BW designates the Baade's 
Window field.}
\end{figure} 

\section{Some results and future plans}  

The first results from our survey were presented by
\citet{ptsw01}, who obtained 2dF spectra at 2.4 \AA\ 
resolution of 130 stars distributed all across the
bulge CMD.  47 stars proved to be hot, and 37 were
BHBs.  They derived temperatures and metallicities
for these by comparison to synthetic spectra based
on the \citet{kurucz} models.  Temperatures were
set independently of reddening using the central
profile (but not the core) of the Balmer lines H$\beta$,
H$\gamma$, and H$\delta$ where necessary.  Most of
the BHB stars were metal-poor, but two had abundances
of solar or higher.

In 2001, we returned to the 2dF and obtained spectra
of over 1000 BHB candidates on three lines of
sight (see Fig.\ 2).  The sample selection included
mainly fainter BHBs than previously.
Most of the sample has S/N $> 40$ down to $B = 17.5$.
Analysis of these spectra is underway. A preliminary
comparison of the spectra against empirical templates
in NGC~6752 obtained on the same observing run reveal
a few sdB candidates, some of which may have composite
spectra indicating a cooler companion like many
field sdB stars \citep{sgb00};  at these magnitudes
sdB stars are likely to be foreground objects.
A few metal-rich BHB stars are seen, but most are
metal poor as was found in the earlier survey.  Preliminary
results from the search for RR Lyraes in these fields
show that only a few percent of the BHB stars are
likely variables.

The main result relevant to this
meeting is shown in Figure 3,  which displays 
a color-magnitude diagram (CMD) in $B$ and $B-V$ for 
the field at $(\ell, b) = (+0^\circ, -6^\circ)$. 
The axes are marked  ``instrumental'' because we are 
still working out the final details of the photometric 
calibration at the $\approx 0.1$ mag level.  
The CMD shows the usual features one 
sees on lines of sight toward the bulge (e.g., 
\citet{tern88,kiraga97}), 
namely a prominent giant branch and clump, and 
a foreground sequence of main-sequence dwarfs. In addition,
there is an extended blue HB, with bluer colors than
the foreground sequence and which extends down to
below $B = 20$.   

The faintest stars on the blue HB are as faint as the confirmed 
EHB stars in globular and open clusters  ($M_V \ge +4$).  
In the metal-rich open cluster NGC~6791, for example, 
sdB stars are found at  $B = 17.9$, $B -V = -0.1$.  
The cluster is 4.8 kpc distant ($m - M_V = 13.42$),
with $(B - V) = 0.1$ \citep{cgl99}.  Were the
cluster located at a distance of 8 kpc with  $E(B -V) = 0.5$, 
the sdB stars would be found at $B \approx 20.8$, 
$B - V = 0.4$, or  $V = 20.2$.  The CMD in Figure 3 
reaches nearly a magnitude fainter.  

We also note that the bulge HB in
all the currently reduced MOSAIC fields appears to
have a gap around $B = 18$;  there are a large number
of cool BHB stars, many faint and blue stars, and
markedly little in between.  This gap occurs at
$M_V = +2$; similar features are seen in the CMDs of
several globular clusters.

The most important remaining observational goal is
to see whether the faint, blue stars we find on the MOSAIC
CMDs are bulge EHB stars.
We know from our 2dF spectra that the majority of the BHB stars
bluer than the foreground disk and with
$B \leq 18$ are probably in the bulge, though there
are also likely to be halo HB stars at larger distances
and the occasional foreground A-type main-sequence star
as found by \citet{ptsw01}.
We do not currently have spectra for the faintest
candidates on Figure 3 to confirm their status as EHB
stars in the bulge.  While the spectroscopy would
be challenging, one could get a reasonable sample since
the surface density of these faint EHB candidates is 
well suited for the current generation of multifiber 
spectrographs:  in the field shown in Figure 3, 
there are several dozen targets in an area 
approximately $26 \times 26$ arcmin.

\begin{figure}[ht]
\centerline{\includegraphics[width=35pc]{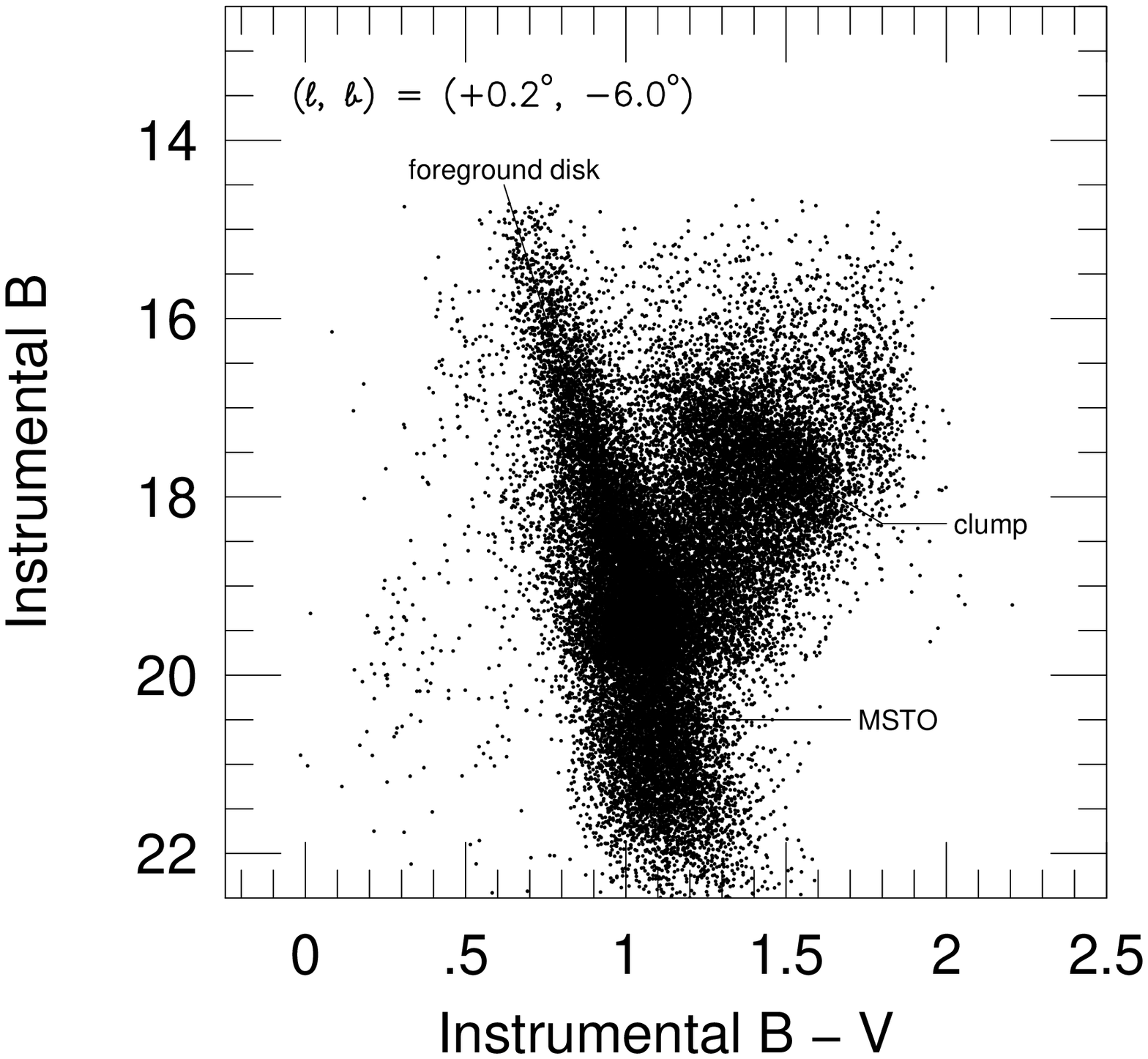}}
\caption{CMD for part of the CTIO MOSAIC field 
at $(\ell, b) = (0^\circ, -6^\circ)$. The total 
area covered by this reduction is 680 square arcmin. 
The zero point and color terms in the reduction 
are not yet finalized.  The HB extends blueward of 
the sequence of foreground stars, turning downward
below $B \approx 17$.  The photometry 
has not been corrected for the (patchy) reddening, 
which averages about $E(B - V) = 0.4$ in this field.
The principal features of the CMD are marked; ``MSTO'' 
indicates the level of the bulge main-sequence turnoff.
Only about 10\% of the stars near the turnoff are
shown in this figure.} 
\end{figure}

\end{document}